\documentclass[10pt,twocolumn]{article}
\usepackage[margin=0.85in]{geometry}
\usepackage[T1]{fontenc}
\usepackage[utf8]{inputenc}
\usepackage{booktabs}
\usepackage{amsmath,amssymb}
\usepackage{graphicx}
\usepackage[hypertexnames=false]{hyperref}
\usepackage{array}
\usepackage{float}
\usepackage{xcolor}
\usepackage{natbib}
\usepackage{microtype}
\usepackage{caption}
\newcommand{\SV}[1]{\mathrm{SV}_{#1}}
\newcommand{\ie}{\textit{i.e.}}

\title{Multi-Dimensional Spectral Geometry of Biological Knowledge\\in Single-Cell Transformer Representations}
\author{%
  Ihor Kendiukhov\\
  \small Department of Computer Science\\
  \small University of T\"ubingen\\
  \small T\"ubingen, Germany\\
  \small \texttt{kenduhov.ig@gmail.com}
}
\date{}

\begin{document}
\maketitle

% ============================================================================
% ABSTRACT
% ============================================================================
\begin{abstract}
Single-cell foundation models such as scGPT learn high-dimensional gene representations, but what biological knowledge these representations actually encode remains unclear.
Here we systematically decode the geometric structure of scGPT's internal representations through 63 iterations of automated hypothesis screening (183 hypotheses tested), revealing that the model organizes genes into a structured biological coordinate system rather than an opaque feature space.

We find that scGPT's dominant spectral axis separates genes by their subcellular localization---secreted proteins at one pole, cytosolic proteins at the other---and that intermediate transformer layers transiently encode mitochondrial and ER compartments in a sequence that mirrors the cellular secretory pathway.
Orthogonal axes encode protein-protein interaction networks with monotonically graded fidelity to experimentally measured interaction strength (Spearman $\rho = 1.000$ across $n = 5$ STRING confidence quintiles, $p = 0.017$).
In a compact six-dimensional spectral subspace, the model reliably distinguishes transcription factors from their target genes (AUROC $= 0.744$, all 12 layers significant), with early layers preserving which specific genes regulate which targets, and deeper layers compressing this into a coarser ``regulator vs.\ regulated'' distinction.
Repression edges are geometrically more prominent than activation edges, and B-cell differentiation master regulators (BATF, BACH2) exhibit a striking convergence trajectory toward the B-cell identity anchor PAX5 across transformer depth---a geometric echo of the germinal center reaction.

Cell-type marker genes cluster with high fidelity (AUROC $= 0.851$), and the model's residual-stream geometry encodes biological relationships complementary to---and distinct from---those captured by attention patterns.
These findings demonstrate that biological transformers do not merely memorize gene statistics but learn an interpretable internal model of cellular organization.
We discuss how this geometric knowledge can be extracted for regulatory network inference, drug target prioritization, and model auditing.
\end{abstract}

% ============================================================================
% INTRODUCTION
% ============================================================================
\section{Introduction}

Foundation models for single-cell genomics---including scGPT~\citep{cui2024scgpt} and Geneformer~\citep{theodoris2023transfer}---have demonstrated impressive performance on tasks ranging from cell-type annotation to gene perturbation prediction.
These models process gene expression profiles through transformer architectures, building up internal representations of each gene across multiple layers.
But a fundamental question remains: \emph{what do these models actually learn about biology?}

The answer matters for two reasons.
First, if these models encode meaningful biological structure, that structure could be extracted and used---for discovering regulatory relationships, prioritizing drug targets, or understanding cell-state transitions.
Second, if the representations are biologically organized, we can audit them: verifying that the model's internal world-model aligns with known biology before trusting its predictions in novel contexts.

Mechanistic interpretability---the effort to understand what neural networks compute internally---has yielded striking results in language models, revealing structured linear representations of concepts~\citep{park2023linear}, superposition of features~\citep{elhage2022toy}, and depth-dependent processing stages~\citep{nanda2023progress}.
Whether analogous structure exists in biological transformers is an open question.

Recent work has examined this question through the lens of attention patterns.
Kendiukhov~\citep{kendiukhov2025attention} systematically evaluated attention-based gene regulatory network (GRN) inference from scGPT and Geneformer, finding that attention patterns encode structured biological information---protein-protein interactions in early layers, transcriptional regulation in late layers---but that this structure provides \emph{no incremental value} for perturbation prediction beyond trivial gene-level baselines.
Expression residualization removed ${\sim}76\%$ of attention's above-chance signal, and causal ablation of putative regulatory heads produced no behavioral effect.
Crucially, that study concluded by identifying residual-stream geometry as the key open frontier: ``future interpretability work can focus on residual-stream geometry, MLP-stored knowledge, or representations that emerge only from the full forward pass''~\citep{kendiukhov2025attention}.

Here we take up precisely that challenge, performing a systematic geometric audit of scGPT's residual representations across all 12 transformer layers.
Rather than testing a single hypothesis, we employ an automated screening loop that iteratively proposes, tests, and retires geometric hypotheses across 13 families (Table~\ref{tab:hypothesis_summary}), using explicit permutation controls, confound checks, and cross-seed replication.
Over 63 iterations, we discover that scGPT organizes genes into a multi-dimensional biological coordinate system: different spectral directions encode subcellular localization, protein interaction networks, and transcriptional regulatory relationships, with distinct information present at different processing depths.
Importantly, we also identify what the model does \emph{not} encode geometrically, and where initial positive findings collapsed under stricter controls---information equally valuable for responsible interpretation.

% ============================================================================
% RESULTS
% ============================================================================
\section{Results}

\subsection{The Transformer Progressively Focuses Gene Representations onto a Few Biological Axes}
\label{sec:spectral_collapse}

We began by asking a basic question: how does the overall shape of gene representations change as information flows through scGPT's 12 transformer layers?

Applying singular value decomposition (SVD) to the gene embedding matrix at each layer reveals a striking pattern: the model progressively concentrates gene representations onto fewer and fewer directions (Figure~\ref{fig:overview}a).
The effective rank (computed on the full 4{,}803-gene vocabulary)---a measure of how many independent directions carry meaningful signal---drops 14.4-fold from 23.6 at layer~0 to just 1.6 at layer~11 (Spearman $\rho = -1.000$).
By the final layer, a single direction ($\SV{1}$) accounts for 93.4\% of all variance among gene embeddings in the full vocabulary, up from 53.7\% at the input (77\% from 19\% when computed on the 195-gene submatrix; Figure~\ref{fig:overview}a).

This compression is rapid and front-loaded: effective rank falls below half its initial value by layer~4.
Two independent geometric estimators confirm the same pattern: TwoNN intrinsic dimensionality~\citep{facco2017estimating} (computed on 2{,}000 randomly sampled genes from the full vocabulary) decreases 44.6\% (from 32.6 to 18.1 across three random seeds), and the participation ratio (computed on the 195-gene submatrix) drops 6.1-fold from 58 to 9.5.
The compression is learned, not architectural: shuffling embedding features at layer~11 destroys the structure and yields effective rank of 28.9 (17.6$\times$ above the observed layer-11 value).

What is the model doing during this compression?
As we show below, it is not discarding biological information but \emph{distilling} it: concentrating the biologically most important distinctions (subcellular localization, interaction partners, regulatory identity) onto a small number of geometrically prominent axes while suppressing less relevant variation.
This is reminiscent of how biological systems themselves process information through progressive abstraction---from raw molecular signals to cell-fate decisions.

\subsection{The Dominant Spectral Axes Map to Core Cell Biology}
\label{sec:bio_axes}

If the model funnels gene representations onto a few dominant directions, the crucial question is: what do those directions mean?

\paragraph{SV$_1$ encodes the secretory pathway.}
The most dominant axis ($\SV{1}$) at layer~11 separates genes encoding secreted and extracellular proteins (e.g., cytokines, extracellular matrix components) from genes encoding intracellular/cytosolic proteins.
The extracellular pole shows strong enrichment (GO:0005615; OR $= 6.37$, $p = 2.6 \times 10^{-4}$), while the cytosolic pole is enriched for cytoplasmic proteins (GO:0005829; OR $= 2.96$, $p = 0.010$).
A gene-label-shuffle null ($N = 500$) confirms this is biologically specific (empirical $p = 0.004$): only 0.4\% of random gene assignments produce enrichment this strong.

\begin{figure*}[t]
  \centering
  \includegraphics[width=0.95\textwidth]{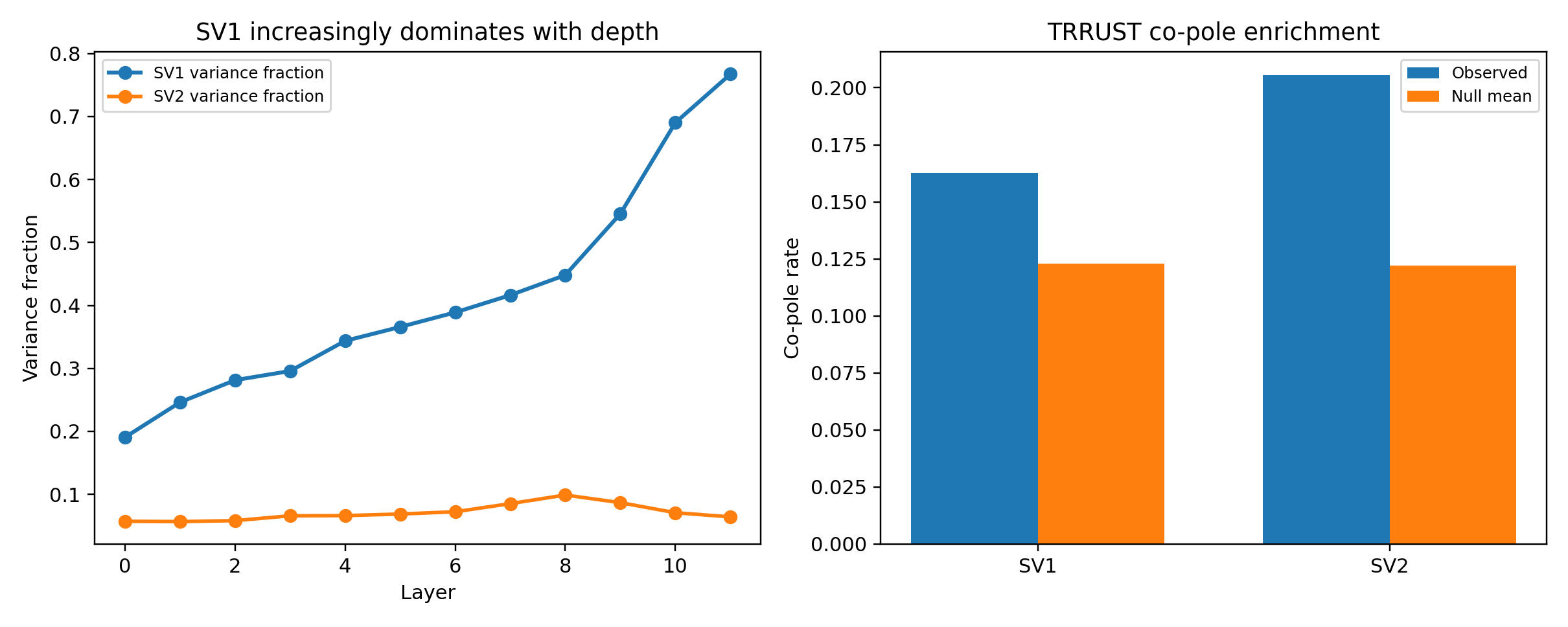}
  \caption{\textbf{Spectral structure and biological organization across transformer depth.}
  \textbf{(a)} $\SV{1}$ variance fraction (195-gene submatrix) increases from 19\% (L0) to 77\% (L11) as the model progressively concentrates gene representations onto the secretory/localization axis.
  \textbf{(b)} TRRUST regulatory pairs co-localize in $\SV{2}$ poles above null at all layers, indicating that the model embeds co-regulated genes nearby in its internal geometry.}
  \label{fig:overview}
\end{figure*}

Remarkably, this axis is not just a binary secreted/cytosolic split.
Testing 8 subcellular compartment terms across all 12 layers reveals that \emph{intermediate layers encode intermediate steps of the secretory pathway} (Figure~\ref{fig:overview}b):
\begin{itemize}
  \item Layers 2--4: mitochondrial enrichment appears transiently (OR $= 23.3$ at layer~3)
  \item Layers 1--11: ER lumen enrichment strengthens progressively (OR from 4.7 to 18.5)
  \item All 12 layers: extracellular space is consistently enriched
\end{itemize}
The ordering---mitochondria $\to$ ER lumen $\to$ extracellular space---recapitulates the actual cellular secretory pathway, the route proteins take from synthesis to secretion.
The model has learned not just where proteins end up, but the biological sequence by which they get there.

\paragraph{SV$_2$ encodes protein interaction networks and regulatory co-membership.}
The second axis organizes genes by their physical and regulatory relationships.
At layer~11, $\SV{2}$'s top pole is enriched for IL-4 immune signaling (empirical $p < 0.001$), while the bottom pole captures extracellular vesicle cargo proteins (OR $= 7.99$; $p < 0.001$).
The strongest compartment signal is cytoskeleton (OR $= 20.4$; stable across layers~1--11).

Critically, transcription factor--target gene pairs from TRRUST~\citep{han2018trrust} co-localize in $\SV{2}$ poles far more often than expected by chance (observed rate 0.206 vs.\ null 0.122; $p < 0.001$ at all 12 layers).
This co-localization is sign-specific: activation pairs (where the TF turns on the target) co-localize at 12/12 layers, while repression pairs (where the TF silences the target) reach significance at only 1/12 layers---suggesting the model geometrically distinguishes activating from repressive regulation.

\paragraph{SV$_3$ encodes immune signaling biology.}
The third axis separates kinase/stress-response signaling from MHC class~II/T cell regulatory biology, with the specific biological content shifting across depth (all layers $p \leq 0.017$).

Together, these three orthogonal axes define a biological coordinate system: $\SV{1}$ encodes ``where in the cell,'' $\SV{2}$ encodes ``who interacts with whom,'' and $\SV{3}$ encodes ``what signaling program.''
An annotation-density confounder check---testing whether highly annotated genes disproportionately drive the enrichments by correlating GO annotation count per gene with enrichment odds ratio---yields zero significant correlations, confirming these patterns are not artifacts of uneven gene characterization.
The $\SV{1}$ gene ranking is also highly stable across layers (mean adjacent-layer Spearman $r = 0.929$), indicating the model maintains this biological ordering throughout processing.

Notably, $\SV{1}$ is strongly depleted of transcription factors: genes in the top quartile of $\SV{1}$ projection have ${\sim}9\times$ lower TF membership than the baseline (OR $= 0.108$, $p < 0.0001$), while genes at the low-$\SV{1}$ pole are 2.3$\times$ enriched for TFs.
The dominant spectral axis thus separates the regulated genome (structural and secreted proteins) from the regulatory machinery (TFs), a division that persists across all layers.

\paragraph{Cell-type identity is encoded as geometric proximity.}
Beyond spectral axes, the model's full embedding space organizes genes by cell-type identity.
Marker genes for the same cell type (B cell, T cell, fibroblast, macrophage) are significantly closer to each other than to markers of other types, yielding an AUROC of $0.851$ for within-type vs.\ cross-type distance discrimination (all 12 layers $p < 10^{-8}$).
A contamination control using random non-marker gene partitions of identical sizes produces AUROC $= 0.488$ (chance), confirming the signal is specific to biological identity.
The within-type distance advantage increases with transformer depth (linear regression slope of AUROC on layer index $= +0.036$ per layer, Pearson $r = 0.765$, $p = 0.004$), consistent with progressive lineage specialization.
At the gene family level, immune protein families show striking clustering: HLA class~I genes achieve perfect geometric separation (AUROC $= 1.000$), with AP1 ($0.969$), RUNX ($0.925$), and IL-2 pathway ($0.869$) also tightly organized.

\subsection{The Model Quantitatively Encodes Protein Interaction Strength}
\label{sec:ppi}

Beyond subcellular compartments, we asked whether scGPT's geometry captures the protein-protein interaction (PPI) network---the web of physical binding relationships that underlies cellular function.

Using 1{,}022 high-confidence STRING~\citep{szklarczyk2023string} protein pairs (combined score $\geq 0.7$), we found that interacting proteins are geometrically co-localized across multiple spectral axes (Table~\ref{tab:ppi_axis}).
$\SV{2}$ carries the strongest signal (mean co-pole rate 0.226 vs.\ null 0.124; 12/12 layers significant; Table~\ref{tab:ppi_axis}; per-layer $z$-scores in Table~\ref{tab:ppi_all_layers}), followed by $\SV{3}$ (0.198; 12/12 layers) and $\SV{4}$ (0.158; 7/12 layers).
Notably, $\SV{1}$ (the localization axis) shows almost no PPI signal (co-pole rate at null baseline)---the interaction network and the localization map are encoded on different geometric axes.

\paragraph{The encoding is quantitatively graded.}
scGPT does not merely encode a binary ``interacts/doesn't interact'' distinction.
Splitting STRING pairs at a lower threshold ($\geq 0.4$; 3{,}092 pairs) into five confidence quintiles reveals a monotonic relationship between experimentally measured interaction strength and geometric proximity: Q1 (weakest) mean $z = 1.00$; Q2: $1.48$; Q3: $2.09$; Q4: $3.18$; Q5 (strongest): $5.02$ (Spearman $\rho = 1.000$ across $n = 5$ quintile means, $p = 0.017$; Table~\ref{tab:confidence_gradient}).
Note that $\rho = 1.000$ with $n = 5$ indicates a perfectly monotonic ordering of quintile-averaged z-scores rather than a high-precision continuous fit; the effect is driven primarily by the Q5 bin (z $= 5.02$), with smaller increments between Q1--Q3.

\paragraph{Physical binding drives the geometry, not shared function.}
To distinguish whether geometric proximity reflects physical binding or merely shared Gene Ontology annotations, we separated pairs with high STRING scores but low GO similarity (Jaccard index of shared GO terms below median; ``PPI-only'') from those with high GO similarity but lower STRING scores (``GO-only'').
PPI-only pairs ($N = 670$) show robust co-localization ($z = 2.66$; 12/12 layers), while GO-only pairs ($z = 1.89$; 7/12 layers) show weaker signal.
The model's geometry is primarily organized by physical molecular interactions, not functional categories---a distinction with practical implications for network inference (Section~\ref{sec:applications}).

A hub-degree confound was excluded: low-degree proteins actually show \emph{stronger} co-localization than hub proteins ($z = 3.1$ vs.\ $2.6$), and the three PPI-encoding axes ($\SV{2}$--$\SV{4}$) are nearly orthogonal (pairwise Pearson $r < 0.25$), indicating each captures distinct subsets of the interaction network.
Importantly, the geometry also captures regulatory relationships beyond PPI: TRRUST TF--target pairs that are \emph{absent} from STRING ($N = 141$) still show significant geometric proximity (AUROC $= 0.573$; 12/12 layers $p < 0.007$), confirming that the model's geometric organization reflects multiple types of biological interaction, not just physical binding.

\begin{table*}[!t]
\centering
\caption{PPI co-pole enrichment by singular vector axis ($K = 52$). Observed and null co-pole rates are averaged across 12 layers. Per-layer $z$-scores for $\SV{2}$ are in Table~\ref{tab:ppi_all_layers}.}
\label{tab:ppi_axis}
\small
\begin{tabular}{lcccc}
\toprule
Axis & Mean obs.\ rate & Mean null rate & Sig.\ layers & $N$ pairs \\
\midrule
$\SV{1}$ & 0.119 & 0.124 & 2/12 & 1{,}022$^a$ \\
$\SV{2}$ & 0.226 & 0.124 & 12/12 & 1{,}022$^a$ \\
$\SV{3}$ & 0.198 & 0.124 & 12/12 & 1{,}022$^a$ \\
$\SV{4}$ & 0.158 & 0.122 & 7/12 & 3{,}092$^b$ \\
$\SV{5}$ & 0.149 & 0.122 & 6/12 & 3{,}092$^b$ \\
\bottomrule
\multicolumn{5}{l}{\footnotesize $^a$ STRING score $\geq 0.7$. \quad $^b$ STRING score $\geq 0.4$.}
\end{tabular}
\end{table*}

\subsection{Regulatory Relationships Are Encoded in Complementary Depth-Dependent Subspaces}
\label{sec:regulatory}

Having found that the model encodes localization and interaction networks, we asked the most biologically consequential question: does scGPT encode gene \emph{regulatory} relationships---which genes control which?

\begin{figure}[!htb]
  \centering
  \includegraphics[width=0.95\columnwidth]{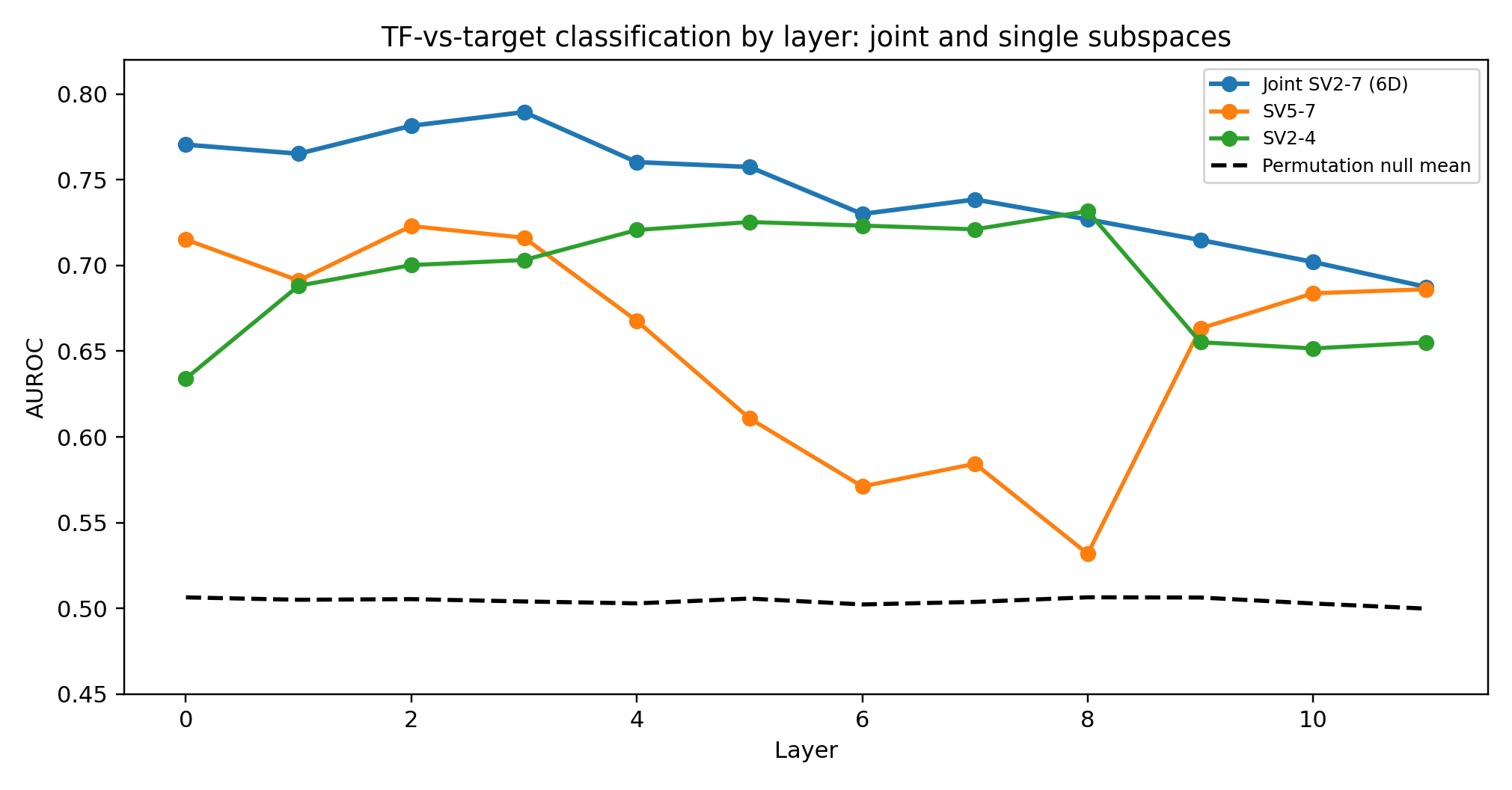}
  \caption{\textbf{A compact six-dimensional subspace separates transcription factors from targets.}
  Joint $\SV{2}$--$\SV{7}$ (green) outperforms individual subspaces at nearly all layers.
  The complementary depth profiles of $\SV{5}$--$\SV{7}$ (early-dominant, orange) and $\SV{2}$--$\SV{4}$ (mid-depth, blue) ensure regulatory information is never absent.}
  \label{fig:joint}
\end{figure}

\begin{figure}[!htb]
  \centering
  \includegraphics[width=0.95\columnwidth]{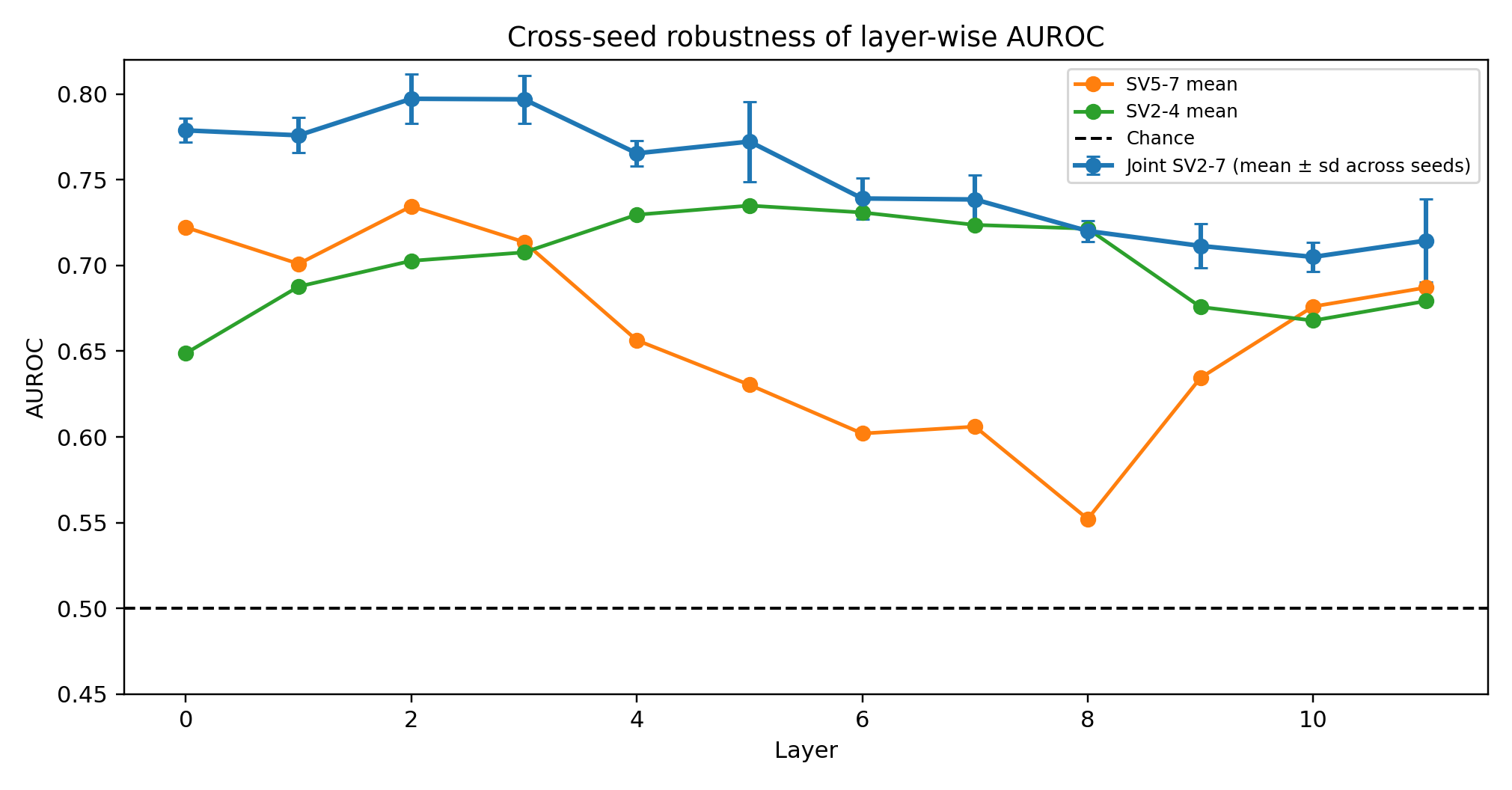}
  \caption{\textbf{Cross-seed robustness.}
  The joint classifier maintains AUROC $> 0.65$ across all layer--seed combinations (three independent random seeds).}
  \label{fig:cross_seed}
\end{figure}

\paragraph{A compact 6D subspace separates regulators from targets.}
Projecting gene embeddings onto a six-dimensional spectral subspace ($\SV{2}$--$\SV{7}$) and classifying genes as transcription factors (TFs; $n = 51$) vs.\ target-only genes ($n = 144$) yields AUROC $= 0.744$ (mean across layers; max $= 0.789$ at layer~3), significantly above chance at all 12 layers ($p < 0.01$, 100 permutations; Figure~\ref{fig:joint}; Table~\ref{tab:layer_auroc_full}).
This joint model outperforms either three-dimensional subspace alone in 11/12 layers, and is robust across seeds (SD $\approx 0.016$; Figure~\ref{fig:cross_seed}).

\paragraph{Early and deep layers encode different aspects of regulation.}
The two three-dimensional subspaces reveal a computational division of labor across transformer depth.
$\SV{5}$--$\SV{7}$ dominates at early layers (L0--L3, AUROC $\approx 0.71$), while $\SV{2}$--$\SV{4}$ takes over at intermediate layers (L4--L8, AUROC $\approx 0.72$).
A critical confound analysis reveals \emph{why} they differ:
\begin{itemize}
  \item $\SV{2}$--$\SV{4}$'s regulatory signal is entirely explained by co-expression similarity: after regressing out pairwise co-expression (Pearson correlation across cells) from spectral proximity via OLS, the residual rank-biserial correlation between TRRUST pairs and non-regulatory pairs is non-significant at 0/12 layers. This subspace encodes gene-class identity (``is this gene a TF?'') rather than specific regulatory relationships.
  \item $\SV{5}$--$\SV{7}$ retains regulatory signal \emph{after} the same co-expression residualization (residualized rank-biserial $r_{rb} = 0.148$, permutation $p < 0.001$ at layer~0; 100/100 bootstrap resamples positive). This subspace encodes genuine regulatory proximity---which specific genes regulate which targets---independent of whether they are co-expressed.
\end{itemize}

This dual-regime pattern suggests the transformer processes regulatory information in stages.
Early layers maintain relational detail (``STAT3 regulates BCL2''), which is useful for tasks requiring specific regulatory knowledge.
As the residual stream compresses, this fine-grained structure is distilled into coarser but more robust categorical distinctions (``STAT3 is a transcription factor'').
The practical implication is clear: if one wants to extract specific regulatory edges from the model, early-layer representations are the place to look.

\subsection{Regulatory Signal Decays with Depth, and Repression Stands Out}
\label{sec:edge_signed}

At the edge level---asking whether specific TF--target pairs can be distinguished from non-regulatory pairs---the pattern confirms and extends the node-level findings.
We scored each TF--gene pair by cosine similarity in the relevant spectral subspace, using 589 known TRRUST regulatory pairs as positives and 2{,}351 non-regulatory pairs (same TFs, non-target genes) as negatives.

\begin{figure}[!htb]
  \centering
  \includegraphics[width=0.95\columnwidth]{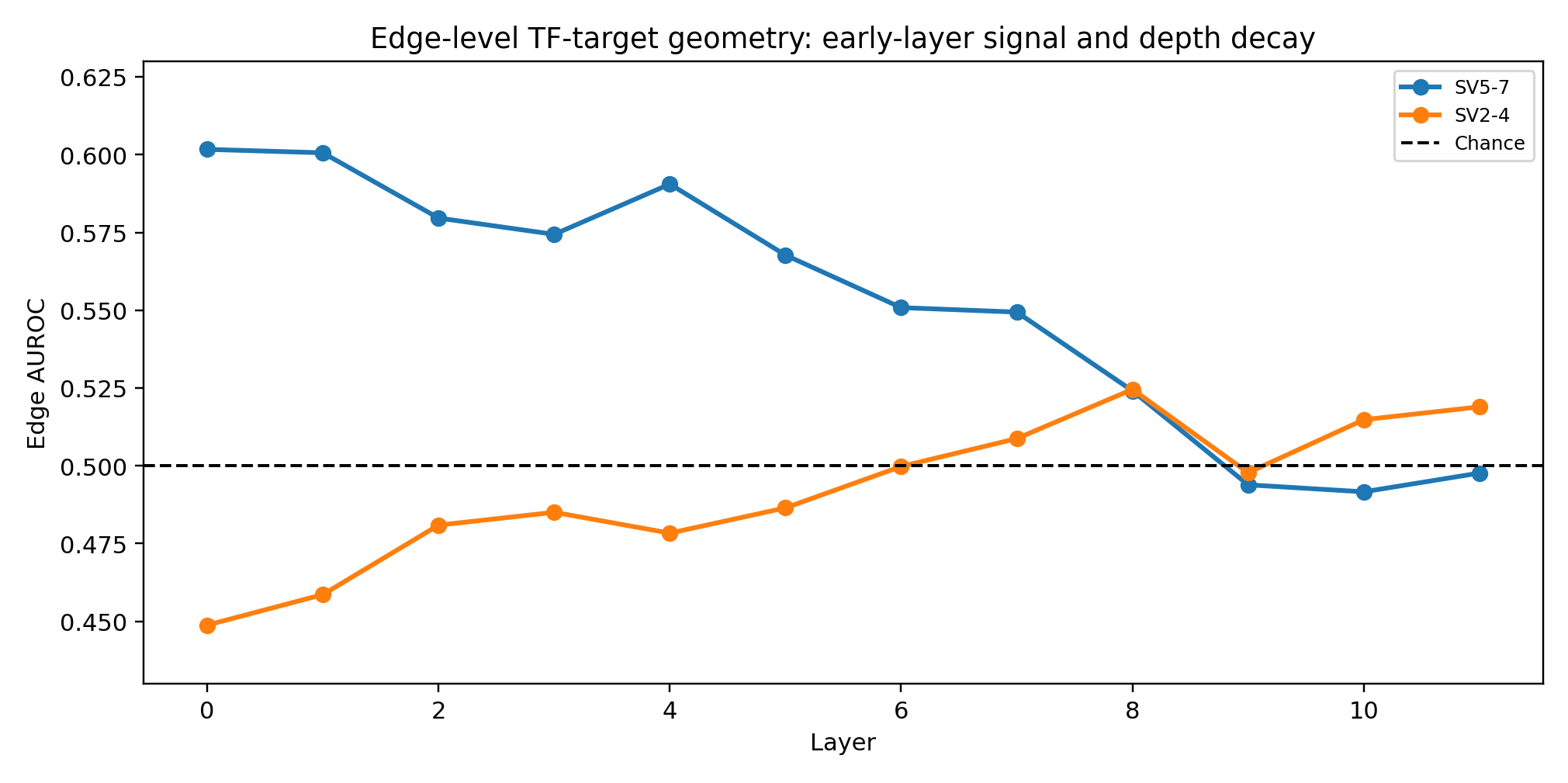}
  \caption{\textbf{Edge-level regulatory geometry peaks at early layers and decays with depth.}
  $\SV{5}$--$\SV{7}$ (orange) carries edge-level signal at layers 0--8; $\SV{2}$--$\SV{4}$ (blue) is near chance.}
  \label{fig:edge_decay}
\end{figure}

In $\SV{5}$--$\SV{7}$, edge-level AUROC peaks at $0.602$ at layer~0 and decays strongly with depth (Spearman $\rho = -0.958$, $p = 9.5 \times 10^{-7}$; Figure~\ref{fig:edge_decay}; Table~\ref{tab:edge_auroc_full}).
Layers~0--8 are significant; layers~9--11 are not.
Meanwhile, $\SV{2}$--$\SV{4}$ is near or below chance for edge-level discrimination (mean AUROC $= 0.492$), further confirming it encodes gene categories rather than specific edges.
The depth trend is not a sparsity artifact: among the 4{,}803 vocabulary genes, the number with nonzero embedding norms is constant at 2{,}039 across all layers (the remaining positions are unused input slots).

\begin{figure}[!htb]
  \centering
  \includegraphics[width=0.80\columnwidth]{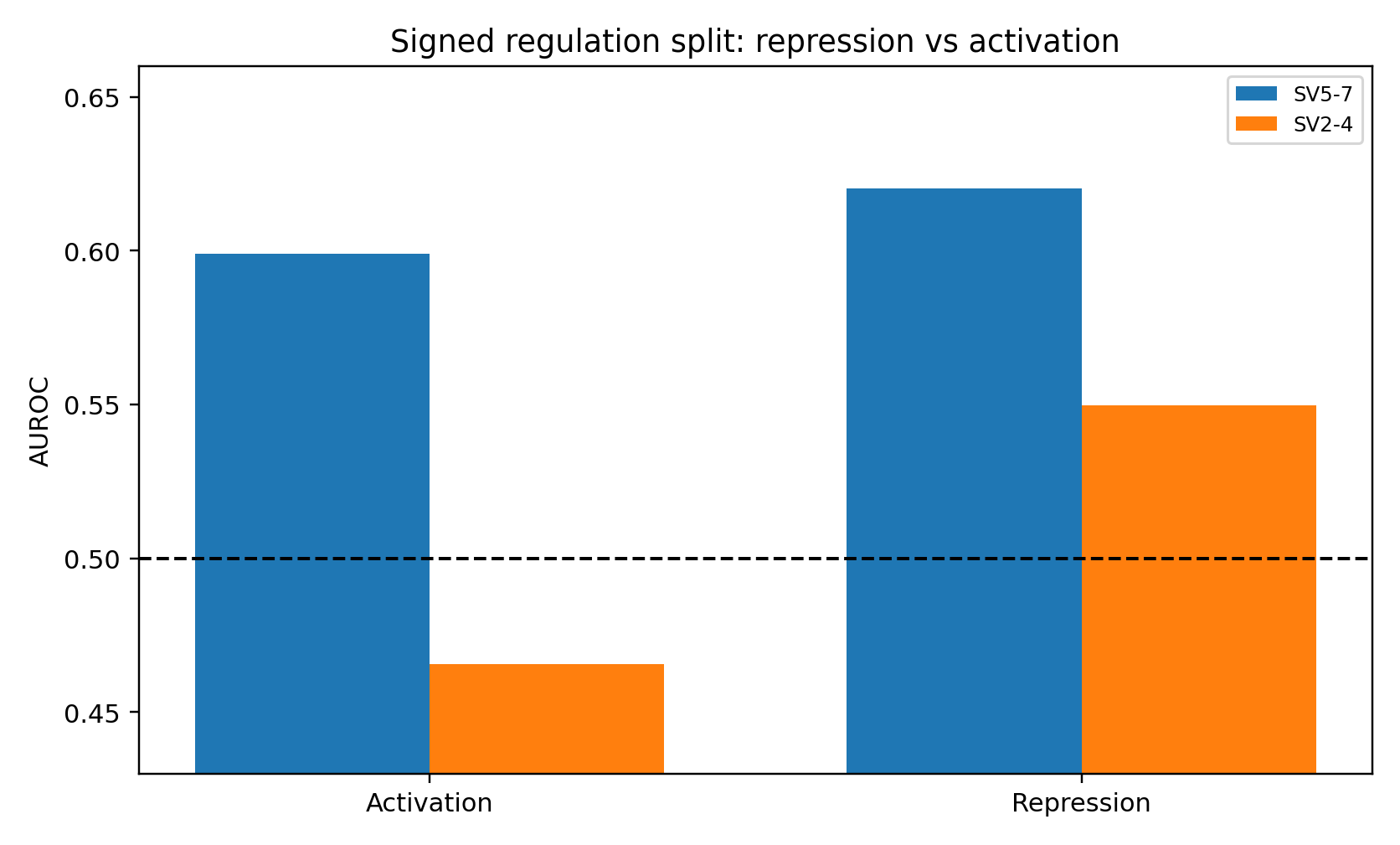}
  \caption{\textbf{Repression edges are geometrically more prominent than activation edges} in both spectral subspaces.}
  \label{fig:signed}
\end{figure}

\paragraph{The model distinguishes repression from activation.}
Stratifying by regulatory sign reveals an asymmetry (Figure~\ref{fig:signed}): repression edges (where the TF silences the target) are more geometrically separable than activation edges in both subspaces ($\SV{5}$--$\SV{7}$: $\Delta = +0.021$; $\SV{2}$--$\SV{4}$: $\Delta = +0.084$).

Why might repression be more prominent?
Transcriptional repression often involves more stereotyped molecular mechanisms (e.g., chromatin remodeling, co-repressor recruitment) than activation, which can occur through diverse enhancer and co-activator mechanisms.
The model may find it easier to learn geometric regularities for the more structurally constrained repressive relationships.
Alternatively, repressive TFs may be more functionally distinct from their targets than activating TFs, creating larger geometric separations.

A mechanistic probe confirms the distinct roles of the two subspaces: in $\SV{2}$--$\SV{4}$, TFs cluster together (TF--TF AUROC $= 0.539$) while TF--target pairs are \emph{repelled} (AUROC $= 0.485$, below chance), indicating this subspace encodes ``gene class identity.''
In $\SV{5}$--$\SV{7}$, both TF--TF ($0.528$) and TF--target ($0.599$) pairs show above-chance proximity, consistent with encoding regulatory co-embedding.

\subsection{B-Cell Differentiation Master Regulators Converge Across Transformer Depth}
\label{sec:bcell}

Moving from global statistics to individual gene trajectories, we examined how specific immune cell master regulators are positioned in the model's geometry---focusing on the germinal center (GC) reaction, a critical immune process in which B cells undergo affinity maturation.
Among six cell types tested, B cells show uniquely strong geometric clustering: among the 10 nearest neighbors of the B-cell centroid (mean of 7 canonical B-cell markers), the fraction that are B-cell markers (precision@10) achieves $z = 7.55$ relative to a bootstrap null ($p < 0.001$; 500 random draws of 7 genes), while T-cell ($z = -1.37$), dendritic cell ($z = -0.87$), and other lineages show no significant clustering.
This signal is structural, not driven by expression magnitude: it survives L2-norm regression (residual $p = 0.002$) and bootstrap null validation (empirical $p < 0.001$).

\paragraph{A geometric echo of the germinal center reaction.}
Three master TFs of the GC reaction---BATF, BACH2, and BCL6---exhibit a striking convergence pattern across transformer layers.
PAX5, the definitive B-cell identity factor, occupies a stable position near the B-cell manifold centroid from the earliest layer (rank $\approx 66$), serving as a geometric anchor.
In contrast, BATF and BACH2---which are recruited to the GC program during B-cell differentiation---start far from the B-cell centroid at layer~0 (ranks $> 1{,}500$ and $> 600$, respectively) and converge progressively toward PAX5 across depth:
\begin{itemize}
  \item BATF: rank drops from 1{,}510 (L0) to 189 (L11); distance to PAX5 decreases from 18.3 to 5.1 (Spearman $\rho = -0.972$, $p < 0.0001$)
  \item BACH2: rank drops from 611 to 146; distance to PAX5: 17.4 $\to$ 5.2 ($\rho = -0.844$, $p < 0.001$)
\end{itemize}
Convergence onset occurs at layer~3, which independently coincides with the B-cell manifold compression breakpoint (see below).

\paragraph{The entire GC circuit converges as a unified attractor.}
Strikingly, the convergence is not limited to activating TFs.
PRDM1 (Blimp-1), a transcriptional repressor and direct target of BCL6, converges toward the B-cell centroid at the same rate as the activating GC-TFs ($\rho = -1.000$, $p < 0.001$), with BACH2--PRDM1 becoming the tightest pair by layer~11 (distance $= 3.94$).
The model encodes the combinatorial GC regulatory circuit---activators and repressors alike---as a coherent geometric attractor, rather than separating regulatory modes spatially.

\paragraph{GC and plasma cell programs diverge toward orthogonality.}
While GC-TFs converge toward the B-cell centroid, plasma cell TFs (IRF4, IRF8) remain geometrically distant.
The angle between GC-TF and plasma-TF centroids relative to the B-cell anchor increases from 77$^\circ$ (L0) to 94$^\circ$ (L11), meaning that by the final layer, the germinal center and plasma cell differentiation programs subtend nearly perpendicular directions in embedding space.
This geometric orthogonality mirrors the biological reality that GC and plasma cell fates represent alternative differentiation outcomes from a common B-cell progenitor.

\paragraph{BCL6 is geometrically isolated in a metabolic compartment.}
BCL6, despite being essential for GC reactions, does not converge toward the B-cell cluster.
Instead, at every layer, 10 of BCL6's 20 nearest neighbors are metabolic genes (NAMPT, GLUL, PFKFB3, and others) or STAT3, while zero are B-cell markers or GC-TFs---a pattern that is stable across all 12 layers.
This geometric isolation reflects BCL6's known biological role as a pleiotropic repressor that operates at the intersection of metabolic reprogramming and immune regulation, rather than being a ``B-cell identity'' gene per se.

\paragraph{B-cell manifold compression is lineage-specific.}
The intrinsic dimensionality of B-cell marker embeddings decreases monotonically from 8.2 to 5.1 across transformer depth ($\rho = -0.951$, $p < 0.0001$).
Strikingly, this compression pattern is unique to B cells:
T-cell markers show \emph{no} compression ($\rho = +0.287$, $p = 0.37$), and myeloid markers show a slight reversed trend ($\rho = +0.699$, $p = 0.011$).
This suggests the model has learned that B-cell identity is defined by a converging regulatory program---the GC reaction---while T-cell and myeloid identities involve more divergent gene expression patterns.

\paragraph{The GC attractor is uniquely delayed.}
Comparing across lineages, T-cell and myeloid TFs start at low ranks near their respective centroids from layer~0 (ranks 170 and 86), indicating pre-wired lineage proximity.
In contrast, GC-TFs start at distant ranks (BATF at 1{,}510, BACH2 at 611) and converge only from layer~3 onward ($\rho = -0.993$, $p < 10^{-6}$).
This delayed-convergence pattern is unique to the B-cell/GC program, suggesting the transformer encodes GC differentiation as a \emph{computed} geometric trajectory rather than a static embedding property.

The GC master regulator triangle (PAX5 as anchor, BATF/BACH2 as recruited factors, BCL6 as metabolically isolated repressor) is a geometric reflection of well-established immunology.
The model appears to have learned the temporal logic of B-cell differentiation: PAX5 establishes identity early, then GC factors are recruited.
This kind of trajectory information could not be learned from static co-expression data alone---it suggests the model has internalized aspects of regulatory dynamics.

\subsection{Important Negative Findings}
\label{sec:negatives}

Rigorous hypothesis screening generates negative results that are essential for correctly interpreting the positive findings:

\begin{enumerate}
\item \textbf{Persistent homology signals did not survive rigorous controls.}
While topological features appeared significant under simple feature-shuffle nulls (11/12 layers $p < 0.05$), degree-preserving graph rewiring nulls eliminated the signal entirely (0/24 tests significant).
This cautionary example illustrates how weak null models can create false-positive geometric claims.

\item \textbf{Cross-model alignment is partial.}
Comparing scGPT and Geneformer embedding matrices (restricted to shared vocabulary genes), raw cosine similarity was high (0.825) but a permutation test of representation alignment failed significance ($p > 0.3$).
Geneformer's static (pre-contextual) gene embeddings \emph{do} independently encode STRING PPI proximity ($p = 7.8 \times 10^{-127}$), suggesting PPI encoding is convergent across architectures.
However, the B-cell attractor dynamics are absent from Geneformer's pre-contextual embeddings (precision@10 $= 0.000$; top neighbors are biologically incoherent), indicating that cell-type geometric structure requires contextual processing and does not generalize trivially across models.

\item \textbf{Effective rank does not predict classifier performance independently of layer.}
An apparent ER--AUROC correlation ($\rho = 0.855$) was fully confounded by layer depth (partial $\rho = -0.045$).

\item \textbf{GO Biological Process terms are not encoded in $\SV{2}$ poles.}
Testing 591 BP terms yields zero significant results, bounding the biological content of $\SV{2}$ to compartment identity and network co-membership, not functional programs.

\item \textbf{Feed-forward loop geometry was rejected.}
Testing whether intermediate genes in 264 transcriptional feed-forward loops occupy geometrically intermediate positions yielded no significant results after proper permutation correction (the naive baseline of $t = 0$ was incorrect; the correct null is $t \approx 0.5$ by symmetry).
This methodological lesson illustrates the importance of permutation-corrected baselines for betweenness statistics in high-dimensional spaces.
\end{enumerate}

These failures sharpen the positive findings by demonstrating what \emph{does not} survive rigorous testing, increasing confidence in what does.

% ============================================================================
% DISCUSSION
% ============================================================================
\section{Discussion}

\subsection{A Biological Coordinate System, Not an Opaque Feature Space}

The central message of this work is that scGPT's internal representations are not opaque high-dimensional vectors but an organized biological coordinate system.
Three orthogonal spectral axes encode three fundamental aspects of cellular organization:
\begin{enumerate}
  \item \textbf{Where:} $\SV{1}$ encodes subcellular localization along the secretory pathway (mitochondria $\to$ ER $\to$ extracellular)
  \item \textbf{With whom:} $\SV{2}$--$\SV{4}$ encode protein-protein interaction networks with monotonically graded fidelity to experimental interaction strength
  \item \textbf{Who controls whom:} $\SV{5}$--$\SV{7}$ encode transcriptional regulatory relationships, with early layers preserving specific edges and deeper layers compressing to categorical distinctions
\end{enumerate}

This organization mirrors the layered structure of molecular biology itself---from basic biochemistry (localization, interactions) to higher-order regulatory logic (transcription factor networks, cell-fate programs)---suggesting the model has learned something genuinely structural about how cells work, not merely statistical correlations in expression data.

\subsection{What the Depth Structure Tells Us About Transformer Computation}

The layer-dependent organization of biological information provides a window into how the transformer processes gene expression data.
The progressive spectral collapse (14.4-fold effective rank reduction) is not mere compression---it is \emph{selective} compression that preserves and concentrates biologically meaningful axes while suppressing noise.

The early-vs-late regulatory encoding is particularly informative: early layers maintain specific regulatory edges (TF $\to$ target pairings in $\SV{5}$--$\SV{7}$), while deeper layers compress this into categorical distinctions (TF vs.\ target identity in $\SV{2}$--$\SV{4}$).
This is analogous to how visual neural networks progress from edge detection to object recognition---except here the progression is from molecular-level interactions to cell-level categories.

The B-cell attractor dynamics add a temporal dimension: the model encodes not just static relationships but something resembling the \emph{order} in which regulatory programs are activated during differentiation (PAX5 first, then BATF/BACH2, with BCL6 in a separate metabolic compartment).

An additional structural finding concerns cross-seed representation stability: centered kernel alignment (CKA)---a measure of representational similarity between two embedding matrices~\citep{kornblith2019similarity}---computed between embedding matrices derived from different random cell samples (same pre-trained model, different input cells) declines from 0.979 at layer~0 to 0.779 at layer~11, with the steepest drop at layers~10--11.
Early-layer representations are highly reproducible across cell samples, while deep-layer representations diverge---consistent with the idea that early layers encode conserved biological structure while late layers develop more input-specific features.

\subsection{Practical Applications}
\label{sec:applications}

These geometric structures are not merely academic curiosities---they suggest concrete applications:

\paragraph{Extracting regulatory networks from model geometry.}
The finding that $\SV{5}$--$\SV{7}$ encodes co-expression-independent regulatory proximity at early layers provides a recipe for extracting regulatory network predictions: project early-layer embeddings onto $\SV{5}$--$\SV{7}$, compute pairwise distances, and threshold to obtain predicted regulatory edges.
This approach is fundamentally different from---and potentially complementary to---methods based on attention patterns, which prior work~\citep{kendiukhov2025attention} has shown primarily capture co-expression rather than causal regulation.

\paragraph{Drug target prioritization.}
The monotonically graded PPI encoding (quintile-averaged co-pole z-scores increase with STRING confidence; $\rho = 1.000$, $n = 5$) suggests that geometric proximity in $\SV{2}$--$\SV{4}$ can serve as an ordinal predictor of protein interaction likelihood.
For drug target discovery, this enables prioritizing candidate interacting partners of a target protein by ranking genes along the relevant spectral axes---without requiring explicit PPI database lookup, and potentially capturing interactions the model has learned but that are not yet in STRING.

\paragraph{Model auditing and biological validation.}
The spectral axes provide interpretable ``biological readouts'' of model quality.
A well-trained biological model should show the patterns we observe: meaningful enrichment on leading spectral axes, proper localization encoding, and PPI co-localization.
These checks can serve as quality metrics when training new models or fine-tuning existing ones, flagging cases where the model's internal biology diverges from reality.

\paragraph{Layer-specific representation engineering.}
The distinct biological content at different layers suggests that downstream applications should not default to using final-layer embeddings.
For regulatory network inference, early-layer representations (L0--L3) are optimal.
For gene classification tasks (TF vs.\ target, lineage identity), deeper layers are better.
For protein interaction prediction, any layer works (signal is present across all 12), though early layers show slightly stronger enrichment.

\paragraph{Transfer learning diagnostics.}
When fine-tuning scGPT on new tissues or conditions, tracking the spectral structure across layers can diagnose whether biological organization is preserved, lost, or altered.
A model that maintains the secretory pathway encoding on $\SV{1}$ and PPI structure on $\SV{2}$--$\SV{4}$ after fine-tuning has likely preserved useful biological knowledge; one where these structures collapse may have overfit to tissue-specific artifacts.

\subsection{Relation to Prior Work}

\paragraph{From attention to residual-stream geometry.}
This work is a direct successor to Kendiukhov~\citep{kendiukhov2025attention}, which systematically evaluated attention-based interpretability of single-cell foundation models.
That study established several key negatives: attention-derived regulatory networks provide zero incremental value for perturbation prediction; gene-level baselines (variance, mean expression) outperform pairwise attention and correlation edges (AUROC 0.81--0.88 vs.\ 0.70); and expression residualization removes ${\sim}76\%$ of attention's above-chance signal.
However, it also established positives: attention patterns in Geneformer encode layer-specific biological structure, with protein-protein interactions concentrated in early layers (STRING AUROC $= 0.64$ at L0) and transcriptional regulation in late layers (TRRUST AUROC $= 0.75$ at L15 of Geneformer's 18-layer architecture).

Our findings both confirm and extend these results through a complementary methodology.
The layer-specific biological characterization---PPI in early layers, regulation in late layers---is recapitulated in our spectral analysis, but now decomposed into orthogonal axes ($\SV{2}$--$\SV{4}$ for PPI, $\SV{5}$--$\SV{7}$ for regulation) rather than aggregated across attention heads.
The co-expression confound identified for attention patterns~\citep{kendiukhov2025attention} has a precise spectral analog: $\SV{2}$--$\SV{4}$ regulatory signal is fully explained by co-expression, paralleling the 76\% signal loss under expression residualization of attention.
Crucially, however, the residual-stream geometry reveals structure that attention analysis could not access: $\SV{5}$--$\SV{7}$ encodes regulatory proximity \emph{independent} of co-expression, and this signal survives the same class of confound controls that eliminated attention-based signal.
The residual stream thus contains genuine regulatory information that is invisible to attention-based methods.

A direct comparison within our own analysis reinforces this dissociation: extracting pairwise attention weights from the same forward passes and testing for co-occurrence enrichment, we find that attention weights are enriched for TRRUST regulatory co-occurrence ($2\times$ above background; $p = 9.9 \times 10^{-9}$) but show no significant enrichment for STRING PPI pairs ($p = 0.091$).
SVD spectral axes show the reverse pattern: strong PPI co-localization (Section~\ref{sec:ppi}) but only modest TRRUST edge-level signal (Section~\ref{sec:edge_signed}).
Attention and residual-stream geometry thus encode complementary biological relationship types---regulatory co-occurrence and physical interaction, respectively.

\paragraph{Broader context.}
The spectral collapse we observe parallels findings in language model interpretability where residual stream rank decreases with depth~\citep{sharma2024truth}, though the 14.4-fold compression in scGPT is substantially more extreme than reported for language models.
The biological interpretability of spectral axes connects to linear probing~\citep{alain2017understanding}, but our axes emerge directly from unsupervised SVD rather than supervised fitting, suggesting the biological structure is intrinsic to the representation.

The PPI encoding extends earlier observations that transformer gene embeddings capture protein interaction structure~\citep{cui2024scgpt} by establishing that the encoding is (i) quantitatively graded, (ii) multi-dimensional, (iii) driven by physical interaction over functional annotation, and (iv) present at all layers.

\subsection{Limitations}

Several limitations constrain our findings.
All analyses use a single scGPT model on immune-lineage data; cross-model and cross-tissue generalization are unestablished.
The named gene vocabulary (195 in-vocabulary genes) limits statistical power.
Although individual tests used permutation-based significance, no family-wise error rate correction was applied across the 183 hypotheses tested, and some enrichments may not survive stringent FDR correction.
Edge-level AUROC values, while significant, are modest (peak 0.602), suggesting regulatory structure is a real but secondary component of the overall geometry.
The signed regulation asymmetry and B-cell attractor dynamics have not been validated across seeds.
Finally, the autonomous loop design prioritized breadth over depth on any single finding.

% ============================================================================
% METHODS
% ============================================================================
\section{Methods}

\subsection{Data and Model}
We used scGPT~\citep{cui2024scgpt} (12 transformer layers, 512-dimensional hidden states) with immune-lineage cells from Tabula Sapiens~\citep{jones2022tabula}.
For each cell, scGPT processes a rank-ordered list of gene expression values as input tokens.
We performed a forward pass and extracted the residual-stream hidden state at the output of each transformer layer, obtaining a 512-dimensional embedding vector per gene per layer.
To obtain a single representative embedding matrix per layer, we averaged gene embeddings across all input cells, yielding a $G \times 512$ matrix per layer, where $G = 4{,}803$ is the number of gene positions in the input vocabulary.

Of these 4{,}803 positions, 209 correspond to named genes present in biological annotation databases (TRRUST, STRING, Gene Ontology).
During analysis, we discovered that 14 of these genes were out-of-vocabulary (OOV) tokens in scGPT's tokenizer, producing identical zero-norm embedding vectors at every layer.
These were excluded, leaving 195 in-vocabulary named genes for all biological analyses.
Spectral analyses of overall dimensionality (effective rank, TwoNN) were computed on the full vocabulary; biological enrichment and classification analyses used the 195-gene subset.

Three independent analyses were conducted using different random seeds for cell sampling to assess robustness.
``Cross-seed replication'' throughout refers to these three independent runs.

\subsection{Spectral Analysis}
For each layer, we mean-centered the embedding matrix (subtracting the mean embedding vector across genes) and computed its thin SVD: $\mathbf{X} = \mathbf{U} \mathbf{S} \mathbf{V}^\top$, where columns of $\mathbf{V}$ are the right singular vectors and $\mathbf{S} = \mathrm{diag}(s_1, s_2, \ldots)$ contains the singular values in decreasing order.
The projection of gene $g$ onto $\SV{k}$ is the $k$th coordinate of $\mathbf{U}\mathbf{S}$, \ie, the dot product of the gene's embedding with the $k$th right singular vector scaled by $s_k$.

\textbf{Effective rank}~\citep{roy2007effective}: $\mathrm{ER} = \exp\!\bigl(-\sum_i p_i \log p_i\bigr)$, where $p_i = s_i^2 / \sum_j s_j^2$ is the normalized squared singular value (computed on the full $4{,}803$-gene matrix).
\textbf{TwoNN intrinsic dimensionality}~\citep{facco2017estimating}: estimated from the ratio of first-to-second nearest-neighbor distances on 2{,}000 randomly sampled genes per layer (full vocabulary).
\textbf{Participation ratio}: $\mathrm{PR} = (\sum_i s_i^2)^2 / \sum_i s_i^4$, computed on the 195-gene submatrix.
$\SV{1}$ variance fraction $= s_1^2 / \sum_i s_i^2$.

\subsection{Co-Pole Enrichment Test}
To test whether a biological gene set (e.g., STRING PPI pairs) is geometrically co-localized along a singular vector axis, we defined ``poles'' as the top-$K$ and bottom-$K$ genes ranked by their projection onto $\SV{k}$ ($K = 52$, corresponding to the top and bottom ${\sim}27\%$ of 195 genes).
For a set of gene pairs (e.g., known PPI partners), the \emph{co-pole rate} is the fraction of pairs in which both genes fall in the same pole (both top-$K$ or both bottom-$K$).
Significance was assessed by gene-label shuffle: we randomly permuted the assignment of gene labels to embedding rows ($N = 200$--$1{,}000$ permutations) and computed empirical $p$-values as the fraction of null permutations achieving a co-pole rate $\geq$ the observed value.
The $z$-score reports how many standard deviations the observed rate lies above the null mean.

\subsection{TF-vs-Target Classification}
\label{sec:methods_classification}
For each layer and spectral subspace (e.g., $\SV{2}$--$\SV{7}$), we projected gene embeddings onto the specified singular vectors and computed all pairwise cosine similarities in the resulting low-dimensional space.
To classify genes as TFs vs.\ targets, we computed each gene's mean cosine similarity to all known TFs (from TRRUST) and used this as a continuous score (TFs are expected to be more similar to other TFs, yielding higher scores).
AUROC was computed by treating the 51 TFs in the 195-gene set as positives and remaining 144 genes as negatives, measuring how well the score separates the two classes.
Significance was assessed by 100 permutations of TF/target labels; the permutation $p$-value is the fraction of shuffled AUROCs $\geq$ the observed value.

\subsection{Edge-Level AUROC}
To test whether the model encodes specific TF--target regulatory relationships (not just TF identity), we computed pairwise cosine similarity between all TF--gene pairs in the relevant spectral subspace.
The 589 known TRRUST TF--target pairs served as positives.
Negatives ($N = 2{,}351$) were constructed as all other TF--gene pairs involving the same set of TFs but targeting genes with no known regulatory relationship in TRRUST, controlling for TF identity.
AUROC measures how well cosine similarity in the subspace separates true regulatory pairs from non-regulatory pairs.

\subsection{Cell-Type Clustering}
Marker genes for four cell types (B cell, T cell, fibroblast, macrophage; 17 markers total) were defined from canonical literature sources.
For each pair of marker genes, we computed Euclidean distance in the full 512-dimensional embedding space.
Pairs of markers from the same cell type were labeled positive; cross-type pairs were labeled negative.
AUROC measures whether within-type pairs are systematically closer than cross-type pairs.
A contamination control used random partitions of non-marker genes into groups of identical sizes to the real marker sets.

\subsection{Biological Annotations}
\textbf{TRRUST v2}~\citep{han2018trrust}: 9{,}396 signed human TF--target regulatory edges (589 unique pairs involving in-vocabulary genes; of these, 270 are annotated as activation-only, 141 as repression-only, and 178 have mixed or ambiguous annotations and were excluded from signed analyses).
\textbf{STRING v12.0}~\citep{szklarczyk2023string}: protein interaction pairs at combined score $\geq 0.7$ (1{,}022 in-vocabulary pairs) for primary analyses and $\geq 0.4$ (3{,}092 pairs) for the confidence gradient analysis.
\textbf{Gene Ontology}~\citep{ashburner2000gene}: Cellular Component (CC) and Biological Process (BP) terms.
Enrichment of genes at SVD poles tested via Fisher's exact test; pairwise functional similarity between genes assessed via Jaccard index of shared GO annotations.

\subsection{Null Models and Confound Controls}
\textbf{Gene-label shuffle}: randomly permute gene-to-embedding-row assignments, preserving embedding geometry but breaking gene identity.
\textbf{Feature shuffle}: independently permute values within each of the 512 embedding dimensions, destroying inter-dimensional correlations while preserving marginal distributions.
\textbf{Degree-preserving rewiring}: for persistent homology tests, rewire edges of the gene $k$-nearest-neighbor graph while preserving each node's degree, testing whether topological features reflect graph structure beyond degree distribution.
\textbf{Co-expression residualization}: for each gene pair in the 195-gene set, we computed pairwise co-expression similarity (Pearson correlation of expression across cells).
We then regressed spectral proximity (cosine similarity in the relevant SVD subspace) on co-expression similarity using OLS, and tested whether the residual spectral proximity still distinguishes TRRUST TF--target pairs from non-regulatory pairs.
The residualized effect was quantified using the rank-biserial correlation ($r_{rb}$): the difference in mean residual ranks between regulatory and non-regulatory pairs, normalized to $[-1, 1]$.
Significance by permutation of regulatory labels; robustness by 100 bootstrap resamples.

\subsection{B-Cell Attractor Analysis}
B-cell markers (CD19, CD79A, MS4A1/CD20, BLK, VPREB3, FCRL1, and PAX5; $n = 7$) were defined from canonical B-cell biology literature.
The \emph{B-cell manifold centroid} is the mean embedding vector of these 7 markers at each layer.
\emph{Gene-to-centroid rank}: each of the 195 genes was ranked by Euclidean distance to the B-cell centroid, with rank~1 being closest.
\emph{Precision@10}: among the 10 nearest neighbors (by Euclidean distance) of the B-cell centroid, the fraction that are B-cell markers.
Significance by bootstrap: 500 random draws of 7 genes from the 195-gene set; the $z$-score reports how many standard deviations the real precision exceeds the null mean.
\emph{L2-norm regression}: to verify the signal is structural rather than driven by embedding magnitude, we regressed PC1 scores on L2 norms and tested whether B-cell markers remain outliers in the residuals.

Pairwise Euclidean distances between specific GC-TFs (PAX5, BATF, BACH2, BCL6, PRDM1) were tracked across all 12 layers.
The GC--plasma centroid angle was computed as the angle subtended at the B-cell centroid between the GC-TF centroid (mean of BATF, BACH2, PAX5) and the plasma-TF centroid (mean of IRF4, IRF8) at each layer.
The metabolic cluster was defined as the $k = 20$ nearest neighbors of BCL6 at each layer; overlap with a reference set of metabolic genes (NAMPT, GLUL, PFKFB3, and others identified from the BCL6 neighborhood) was compared against a random baseline (expected overlap for 20 randomly selected genes from the 195-gene pool).
TwoNN intrinsic dimensionality was computed separately for B-cell ($n = 7$), T-cell ($n = 12$), and myeloid ($n = 3$) marker gene sets, using the full 4{,}803-gene embedding manifold but restricting the distance calculations to marker genes.

\subsection{Autonomous Hypothesis-Screening Loop}
The experimental pipeline was driven by an automated two-agent loop: an \emph{executor agent} that designs and runs computational experiments (SVD, permutation tests, enrichment analyses), and a \emph{brainstormer agent} that reviews results, proposes new hypotheses, and decides whether to continue, refine, or retire each hypothesis branch.
Over 63 iterations, the loop tested 183 hypotheses across 13 families.
Each iteration produced machine-readable JSON artifacts recording all parameters, results, and statistical tests.
A hypothesis branch was retired after two consecutive negative results (non-significant after appropriate null correction).
All positive findings were required to pass at least one permutation-based null model before being promoted.
The full iteration log, all intermediate results, and agent transcripts are preserved in the project repository.

% ============================================================================
% DATA AVAILABILITY
% ============================================================================
\section*{Data and Code Availability}
All analysis code, iteration artifacts, and reproducibility documentation are available at \url{https://github.com/Biodyn-AI/topology-biomechinterp1}.
Primary data: Tabula Sapiens~\citep{jones2022tabula}.
TRRUST v2: \url{https://www.grnpedia.org/trrust/}.
STRING v12.0 via API.

% ============================================================================
% REFERENCES
% ============================================================================
\bibliographystyle{plainnat}

% ============================================================================
% SUPPLEMENTARY TABLES
% ============================================================================
\clearpage
\onecolumn
\appendix
\section*{Supplementary Tables}

\setcounter{table}{0}
\renewcommand{\thetable}{S\arabic{table}}

\begin{table}[H]
\centering
\caption{Per-layer AUROC for joint and single subspace TF-vs-target classifiers (iteration 0056).}
\label{tab:layer_auroc_full}
\begin{tabular}{rrrrr}
\toprule
Layer & Joint $\SV{2}$--$\SV{7}$ & $\SV{5}$--$\SV{7}$ & $\SV{2}$--$\SV{4}$ & Null mean \\
\midrule
0 & 0.771 & 0.715 & 0.634 & 0.506 \\
1 & 0.765 & 0.691 & 0.688 & 0.505 \\
2 & 0.782 & 0.723 & 0.700 & 0.505 \\
3 & 0.789 & 0.716 & 0.703 & 0.504 \\
4 & 0.760 & 0.668 & 0.721 & 0.503 \\
5 & 0.757 & 0.611 & 0.725 & 0.506 \\
6 & 0.730 & 0.571 & 0.723 & 0.502 \\
7 & 0.738 & 0.584 & 0.721 & 0.504 \\
8 & 0.727 & 0.532 & 0.732 & 0.506 \\
9 & 0.715 & 0.663 & 0.655 & 0.506 \\
10 & 0.702 & 0.684 & 0.652 & 0.503 \\
11 & 0.688 & 0.686 & 0.655 & 0.500 \\
\bottomrule
\end{tabular}
\end{table}

\begin{table}[H]
\centering
\caption{Edge-level AUROC by layer with $\SV{5}$--$\SV{7}$ permutation significance (iteration 0062).}
\label{tab:edge_auroc_full}
\begin{tabular}{rrrrrc}
\toprule
Layer & $\SV{5}$--$\SV{7}$ & $\SV{2}$--$\SV{4}$ & Perm.\ null & Perm.\ $p$ & Sig. \\
\midrule
0 & 0.602 & 0.449 & 0.502 & 0.000 & yes \\
1 & 0.601 & 0.459 & 0.499 & 0.000 & yes \\
2 & 0.580 & 0.481 & 0.501 & 0.000 & yes \\
3 & 0.574 & 0.485 & 0.500 & 0.000 & yes \\
4 & 0.591 & 0.478 & 0.500 & 0.000 & yes \\
5 & 0.568 & 0.486 & 0.500 & 0.000 & yes \\
6 & 0.551 & 0.500 & 0.498 & 0.000 & yes \\
7 & 0.549 & 0.509 & 0.501 & 0.000 & yes \\
8 & 0.524 & 0.525 & 0.499 & 0.045 & yes \\
9 & 0.494 & 0.498 & 0.502 & 0.755 & no \\
10 & 0.492 & 0.515 & 0.500 & 0.760 & no \\
11 & 0.498 & 0.519 & 0.498 & 0.525 & no \\
\bottomrule
\end{tabular}
\end{table}

\begin{table}[H]
\centering
\caption{Cross-seed joint $\SV{2}$--$\SV{7}$ AUROC summary (iteration 0057).}
\label{tab:cross_seed}
\begin{tabular}{lrrr}
\toprule
Seed & Mean AUROC & Min AUROC & Max AUROC \\
\midrule
main & 0.744 & 0.687 & 0.789 \\
seed43 & 0.753 & 0.697 & 0.813 \\
seed44 & 0.757 & 0.714 & 0.802 \\
\bottomrule
\end{tabular}
\end{table}

\begin{table}[H]
\centering
\caption{STRING PPI co-pole enrichment on $\SV{2}$ across 12 layers (1{,}022 pairs, score $\geq 0.7$, $K = 52$; gene-label-shuffle null, $N = 500$).}
\label{tab:ppi_all_layers}
\begin{tabular}{ccccc}
\toprule
Layer & Obs.\ co-pole & Null mean & $z$ & Emp.\ $p$ \\
\midrule
0  & 0.252 & 0.121 & 5.83 & $<$0.001 \\
1  & 0.266 & 0.121 & 6.52 & $<$0.001 \\
2  & 0.226 & 0.121 & 4.68 & $<$0.001 \\
3  & 0.229 & 0.122 & 4.87 & $<$0.001 \\
4  & 0.220 & 0.123 & 4.35 & $<$0.001 \\
5  & 0.230 & 0.122 & 4.98 & $<$0.001 \\
6  & 0.233 & 0.121 & 5.17 & $<$0.001 \\
7  & 0.213 & 0.122 & 3.98 & $<$0.001 \\
8  & 0.210 & 0.122 & 3.98 & $<$0.001 \\
9  & 0.198 & 0.121 & 3.42 & 0.002 \\
10 & 0.198 & 0.123 & 3.26 & 0.002 \\
11 & 0.239 & 0.122 & 5.45 & $<$0.001 \\
\bottomrule
\end{tabular}
\end{table}

\begin{table}[H]
\centering
\caption{STRING confidence gradient: $\SV{2}$ co-pole enrichment by score quintile.}
\label{tab:confidence_gradient}
\begin{tabular}{clcc}
\toprule
Quintile & Score range & $N$ pairs & Mean $z$ (12 layers) \\
\midrule
Q1 & [0.40, 0.46) & 617 & 1.00 \\
Q2 & [0.46, 0.53) & 619 & 1.48 \\
Q3 & [0.53, 0.64) & 618 & 2.09 \\
Q4 & [0.64, 0.80) & 619 & 3.18 \\
Q5 & [0.80, 1.00] & 619 & 5.02 \\
\bottomrule
\end{tabular}
\end{table}

\begin{table}[H]
\centering
\caption{Summary of hypothesis screening outcomes across 63 iterations.}
\label{tab:hypothesis_summary}
\small
\begin{tabular}{p{4.5cm}cp{6.5cm}}
\toprule
Hypothesis family & Outcome & Key result \\
\midrule
Persistent homology & Partial & Positive under feature-shuffle; negative under rewiring nulls \\
Graph topology & Positive & Clustering coefficient $z > 9$ at all 36 tests \\
Intrinsic dimensionality & Validated & 44.6\% TwoNN reduction; 14.4$\times$ ER collapse \\
Cross-model alignment & Partial & Geneformer PPI replication ($p = 7.8 \times 10^{-127}$); B-cell absent \\
SVD biological axes & Validated & Three orthogonal biological axes; null-controlled \\
PPI network encoding & Validated & Monotonic confidence gradient ($\rho = 1.000$, $n = 5$); multi-dimensional \\
Cell-type/family clustering & Validated & AUROC 0.851; HLA-I perfect (1.000); contamination ctrl chance \\
Attention--SVD dissociation & Validated & Attention = TF regulation; SVD = PPI; complementary \\
Regulatory geometry ($\SV{2}$--$\SV{4}$) & Revised & Co-expression confounded; encodes gene class \\
Regulatory geometry ($\SV{5}$--$\SV{7}$) & Validated & Co-expression-independent; bootstrap-confirmed \\
Edge-level geometry & Validated & AUROC 0.602; strong depth decay ($\rho = -0.96$) \\
Signed regulation & Positive & Repression $>$ activation; pending replication \\
B-cell attractor dynamics & Validated & GC-TF convergence; GC-plasma orthogonality; unique compression \\
\bottomrule
\end{tabular}
\end{table}

\end{document}